CURRENT-DENSITY FUNCTIONAL THEORY OF TIME-DEPENDENT LINEAR
RESPONSE IN QUANTAL FLUIDS: RECENT PROGRESS

M. P. TOSI, M. L. CHIOFALO, A. MINGUZZI and R. NIFOSÌ
*Istituto Nazionale di Fisica della Materia and Classe di Scienze*
*Scuola Normale Superiore, I-56126 Pisa, Italy*

**Abstract.** Vignale and Kohn have recently formulated a local density approximation to the time-dependent linear response of an inhomogeneous electron system in terms of a vector potential for exchange and correlation. The vector potential depends on the induced current density through spectral kernels to be evaluated on the homogeneous electron gas. After a brief review of their theory, the case of inhomogeneous Bose superfluids is considered, with main focus on dynamic Kohn-Sham equations for the condensate in the linear response regime and on quantal generalized hydrodynamic equations in the weak inhomogeneity limit. We also present the results of calculations of the exchange-correlation spectra in both electron and superfluid boson systems.

## 1. Introduction

The evaluation of the dynamical properties of inhomogeneous quantal systems is a central problem in condensed matter physics. As an example we may recall the current efforts at elucidating the dynamical behaviour of dilute alkali vapours which have undergone Bose-Einstein condensation in magnetic traps. The experimental studies have concerned (i) the excitation of low-lying shape deformation modes in a regime where there is no detectable noncondensate fraction [1, 2] and the behaviour of mode frequencies and relaxation times as temperature is raised towards and above condensation [3]; (ii) the propagation of sound waves in the condensate and its thermal cloud [4, 5]; and (iii) antisymmetric oscillations of the condensate and the thermal cloud corresponding to second sound [5]. These experiments have stimulated a number of theoretical studies (see *e.g.* [6] and references given therein). The relationship between the dynamics of such confined fluids and that of a homogeneous Bose superfluid [7, 8] also is of great interest.

Local density approximations have been very useful in accounting for exchange and correlation (xc) in the ground state energy of inhomogeneous electron systems, and the search for dynamic extensions has attracted interest for some time (see *e.g.* [9]). A *scalar* time-dependent xc potential, which is a local functional of the time-dependent particle density, is useful in dealing with low-frequency phenomena. However, analysis of the constraints coming from basic conservation laws has shown that inconsistencies can arise and are associated with the non-existence of a gradient expansion at finite frequency ω for the xc potential in terms of the density alone (see *e.g.* [10]). Recently Vignale and Kohn [11, 12] have overcome these difficulties by resorting to a dynamic xc *vector* potential even in the case of a system subject to an external *scalar* potential. They got an expression for the xc vector potential in the linear response regime in terms of kernels to be evaluated on the homogeneous electron gas at the local equilibrium density. Their expression becomes exact when the equilibrium density and the external potential are slowly varying in space, on length scales set by $1/k_F$ and $v_F/\omega$ where $k_F$ and $v_F$ are the local Fermi wave-number and velocity. The kernels are directly related to current-current response functions [13].

In § 2 below we give a brief review of the work of Vignale and Kohn and then proceed in § 3 to report on our work regarding an inhomogeneous superfluid. Again for the latter the validity of the results is restricted to weak inhomogeneity (slow variations over length scales set by the interatomic distance and by $c/\omega$ where c is the local speed of sound) in the linear response regime. In § 4 we report some examples of xc spectra for both the electron fluid and the Bose superfluid, which have been obtained in basically the same decoupling scheme for the current-current correlations [13].

## 2. Weakly Inhomogeneous Electron Fluid in the Normal State

According to the Runge-Gross theorem [10] the problem of a many-electron system in an external potential $V(\mathbf{r},t)$ can be mapped into that of non-interacting electrons in an effective potential $V(\mathbf{r},t) + V_{ex}(\mathbf{r},t)$ (the excess potential $V_{ex}(\mathbf{r},t)$ includes both the Hartree term and the xc term). In the so-called "adiabatic local density approximation" (ALDA) the dynamic xc potential is evaluated from the xc energy density of the homogeneous electron gas in the form

$$V_{xc}^{ALDA}(\mathbf{r},t) = d\varepsilon_{xc}^h(n(\mathbf{r},t))/dn \quad , \tag{1}$$

as in the corresponding static problem. The ALDA embodies the correct electron-gas compressibility, but does not account correctly for plasmon dipersion and omits plasmon damping as well as transverse-current fluctuations. In the context of linearized hydrodynamics for a monatomic fluid, the ALDA embodies the speed of isothermal first sound, but does not account for sound-wave damping nor does it allow for the transition from a collisional to a collisionless regime which may take place *e.g.* as temperature is varied.

As already remarked in § 1, Vignale and Kohn [11, 12] have proposed using a dynamic xc vector potential in the linear-response regime to overcome the limitations of the ALDA for the dynamics of an inhomogeneous electron system in the normal state. The induced current density is written as

$$\delta\mathbf{j}(\mathbf{r},\omega) = \int d\mathbf{r}' \breve{\chi}_{KS}(\mathbf{r},\mathbf{r}';\omega)[\mathbf{a}(\mathbf{r},\omega) + \mathbf{a}_H(\mathbf{r},\omega) + \mathbf{a}_{xc}(\mathbf{r},\omega)] \quad , \tag{2}$$

where $\breve{\chi}_{KS}(\mathbf{r},\mathbf{r}';\omega)$ is the Kohn-Sham response of a non-interacting reference system, $\mathbf{a}(\mathbf{r},\omega)$ is the applied potential, $\mathbf{a}_H(\mathbf{r},\omega)$ is the Hartree mean-field term and $\mathbf{a}_{xc}(\mathbf{r},\omega)$ is the xc vector potential. This is related to the induced current density by

$$\mathbf{a}_{xc}(\mathbf{r},\omega) = \int d\mathbf{r}' \breve{f}_{xc}(\mathbf{r},\mathbf{r}';\omega)\delta\mathbf{j}(\mathbf{r}',\omega) \quad . \tag{3}$$

The kernels $\breve{f}_{xc}(\mathbf{r},\mathbf{r}';\omega)$ entering Eq. (3) have some general exact properties, which are needed to relate them in the weak-inhomogeneity limit to analogous kernels for the homogeneous electron fluid taken at the local equilibrium electron density $n(\mathbf{r})$ [12]. These are Onsager's reciprocity relation, the zero-force and zero-torque theorems (in brief, the net force and the net torque exerted by the xc potential on the system must vanish) and a Ward identity embodying the compressibility sum rule for the homogeneous electron fluid. As the final result the dynamic correction to the ALDA xc vector potential is expressed in terms of the induced *velocity* $\mathbf{u}(\mathbf{r},\omega) = \delta \mathbf{j}(\mathbf{r},\omega)/n(\mathbf{r})$ and of the homogeneous xc current-density kernel $\breve{f}_{xc}(\mathbf{k};\omega)$, evaluated at the local equilibrium density. The latter may be separated into the sum of a longitudinal and a transverse part:

$$\breve{f}_{xc}^h(\mathbf{k};\omega) = \omega^{-2}[f_{xc}^L(\omega)\mathbf{kk} + f_{xc}^T(\omega)(k^2\breve{\mathbf{I}} - \mathbf{kk})] \quad . \tag{4}$$

Rather than reporting the complicated expression of the xc vector potential derived by Vignale and Kohn [11] in the local density approximation, we refer to the later work of Vignale *et al.* [14], who rewrite it in a physically transparent way through a visco-elastic stress tensor $\breve{\sigma}$. Defining

$$\sigma_{ij}^{(xc)} = \eta_{xc}(\omega)\left(\frac{\partial u_i}{\partial r_j} + \frac{\partial u_j}{\partial r_i} - \frac{2}{3}\delta_{ij}\vec{\nabla}\cdot\mathbf{u}\right) + \zeta_{xc}(\omega)\delta_{ij}\vec{\nabla}\cdot\mathbf{u} \tag{5}$$

where $\eta_{xc}(\omega)$ and $\zeta_{xc}(\omega)$ are complex viscosity coefficients given by

$$\eta_{xc}(\omega;n) = -(n^2/i\omega)f_{xc}^T(\omega;n) \tag{6}$$

and

$$\zeta_{xc}(\omega;n) = -(n^2/i\omega)[f_{xc}^L(\omega;n) - \frac{4}{3}f_{xc}^T(\omega;n) - d^2\varepsilon_{xc}(n)/dn^2] \quad , \tag{7}$$

one finds that the divergence of $\breve{\sigma}^{(xc)}$ determines the dynamical xc vector potential. The connection of the theory with generalized hydrodynamics in the linear-response, long-wavelength regime becomes fully evident when one introduces a weak-inhomogeneity assumption also for the Kohn-Sham response matrix, by replacing it by its diamagnetic part $\breve{\chi}_{KS}(\mathbf{r},\mathbf{r}';\omega) \rightarrow \delta(\mathbf{r}-\mathbf{r}')n(\mathbf{r},\omega)/m$. The induced current is then given by

$$-im\omega\delta\mathbf{j}(\mathbf{r},\omega) = n(\mathbf{r})[-i\omega\mathbf{a}(\mathbf{r},\omega) - \vec{\nabla}\delta v_H(\mathbf{r},\omega) - \vec{\nabla}(\delta n(\mathbf{r},\omega)/n^2K_T)] + \vec{\nabla}\cdot\breve{\sigma}(\mathbf{r},\omega) \tag{8}$$

where $\delta v_H(\mathbf{r},\omega)$ is the Hartree potential associated with the long-range Coulomb interactions, $K_T$ is the local electron-gas compressibility and the full stress tensor $\breve{\sigma}$ differs from the expression in Eq. (5) by the replacement $\eta_{xc}(\omega) \rightarrow \eta_{xc}(\omega) - p_0(n)/i\omega$, $p_0$ being the pressure of the non-interacting electron gas.

In summary, the corrections to the ALDA restoring force in the linear-response, weak-inhomogeneity regime merely involve the visco-elastic effects associated with frequency-dependent elastic constants and damping coefficients, which are related to the imaginary part and the real part of the two viscosity coefficients $\eta_{xc}(\omega)$ and $\zeta_{xc}(\omega)$, respectively. For longitudinal motions, in particular, the plasmon excitation arises from the Hartree term in Eq. (8) and the role of the visco-elastic function $\zeta_{xc}(\omega) + 4\eta_{xc}(\omega)/3$ is (i) to shift the plasmon dispersion coefficient away from the low-frequency value determined by $K_T$ in Eq. (8), and (ii) to introduce plasmon damping from decay into multiple electron-hole pairs.

Equation (8) relates the generalized viscosity coefficients $\eta_{xc}(\omega)$ and $\zeta_{xc}(\omega)$, or equivalently the dynamic local field factors $f_{xc}^L(\omega)$ and $f_{xc}^T(\omega)$ entering Eqs (6) and (7), to the current-current response function of the homogeneous electron gas, *via* a finite-frequency generalization of the Kubo relations [13]. This property will be used in § 4 for an evaluation of these xc spectra, based on the exact equation of motion for the current response and on the reduction to two-pair excitation processes through a decoupling approximation [13, 15]. We shall first turn, however, to a discussion of an inhomogeneous Bose superfluid.

## 3. Dynamic Density Functional Theory for a Bose Superfluid

In a superfluid of Bose particles the set of basic dynamic variables must include the superfluid velocity field $\mathbf{u}_s(\mathbf{r},t)$ in addition to the total (superfluid plus normal-fluid) current density. We follow Hohenberg and Martin [7] in introducing the transformation

$$\psi^\dagger(\mathbf{r}) = [\hat{n}_c(\mathbf{r})]^{1/2} \exp[-i\hat{\varphi}(\mathbf{r})] \tag{9}$$

for the field operator in terms of the condensate density and phase operators: then the superfluid velocity operator is $\hat{\mathbf{u}}_s(\mathbf{r}) = m^{-1}\vec{\nabla}\hat{\varphi}(\mathbf{r})$, in accord with the irrotational character taken by the superfluid velocity field in the absence of vortices.

The linear response of the condensate density and phase to a dynamic gauge-breaking external field can be treated by means of the equation of motion for the field operator and is found to have the Hohenberg-Kohn-Sham structure compatible with a mapping of the interacting system into a single-particle reference system [16]. The condensate self-energy $\sigma(1)$ is introduced by

$$\sigma(1) = \int d\bar{2}\, v(\mathbf{r}_1,\mathbf{r}_{\bar{2}}) <\psi^\dagger(\bar{2})\psi(\bar{2})\psi(1)> \tag{10}$$

where $1 = (\mathbf{r}_1,t_1)$, $v(\mathbf{r}_1,\mathbf{r}_{\bar{2}})$ is the interparticle pair potential and the symbol $<\cdots>$ denotes the expectation value of an operator on the equilibrium ensemble at given temperature. The Kohn-Sham response functions $\chi_{KS}(1,1')$ of the condensate are constructed from a reference Green's function $G_{KS}(1,1')$ which is given by

$$G_{KS}^{-1}(1,1') = [i\frac{\partial}{\partial t_1} + \frac{1}{2m}\nabla^2 - V(\mathbf{r}_1) + \mu - \frac{\sigma_{eq}(\mathbf{r}_1)}{<\psi(\mathbf{r}_1)>_{eq}}]\delta(1,1') \quad, \tag{11}$$

where $V(\mathbf{r}_1)$ is the confining potential and $\mu$ is the chemical potential. The two-by-two matrix $\chi(1,1')$ of the response functions of the condensate can then be cast into the form

$$\chi = \chi_{KS} + \chi_{KS} \otimes K \otimes \chi \tag{12}$$

where the symbol $\otimes$ denotes integration over intermediate variables and the kernels in the matrix $K$ are obtained as functional derivatives of the real and imaginary parts of

$$\delta\sigma(1) = \sigma(1) - \sigma_{eq}(\mathbf{r}_1)<\psi(1)>/<\psi(\mathbf{r}_1)>_{eq} \tag{13}$$

with respect to the condensate density and phase.

We should emphasize that (i) the Kohn-Sham reference system that we are proposing for the superfluid contains the interactions through the equilibrium value of the condensate self-energy entering the RHS of Eq. (11); and (ii) we are *not* advocating a density functional approach to the evaluation of the equilibrium state of the superfluid. In particular, a thermodynamic treatment based on the Gross-Pitaevskii equation already is in quantitative agreement with experiment for confined vapours of alkali atoms which have undergone Bose-Einstein condensation [17].

A similar fully microscopic treatment of the current density induced in the inhomogeneous superfluid by a further external driving field leads to microscopic expressions for the remaining response functions (the current-current response and the current-condensate cross response) [16]. While these expressions are quite complicated in the general case, these calculations allow one to complete contact with the well known two-fluid model of Tisza and Landau. Firstly, the total current-current response function has the general structure already found by Hohenberg and Martin [7], who used it to show that in an appropriate limit this function gives the response of the total current density to an external field defining the normal fluid velocity $\mathbf{u}_n(\mathbf{r},t)$. Secondly, a Ward identity for the current response of the noncondensate at fixed condensate, which was first given by Huang and Klein [18] can be used to introduce a function $\rho_s(1,2)$ through the equation

$$\vec{\nabla}_1 \rho_s(1,2) = \vec{\nabla}_1 [n_c(1)\delta(1,2)] - 2m \, \text{Im}[<\psi^\dagger(1)> \frac{\delta\sigma(1)}{\delta\mathbf{a}(2)}|_{<\psi>}] \quad . \tag{14}$$

Within our microscopic treatment Eq. (14) can be rewritten in the form

$$\vec{\nabla}_1 \rho_s(1,2) = \vec{\nabla}_2 \cdot \frac{\delta\mathbf{j}(1)}{\delta\mathbf{u}_s(2)}|_\mathbf{a} \quad . \tag{15}$$

It follows from Eq. (15) that the function $\rho_s(1,2)$ reduces for the homogeneous fluid in the static limit to the hydrodynamic definition of superfluid density given by Hohenberg and Martin [7]. It can also be shown that in the same limit the definition of superfluid density given by Griffin [8] is recovered from Eq. (14).

On the above grounds we have developed a two-fluid model for the generalized hydrodynamics of the inhomogeneous superfluid in the linear-response, weak-inhomogeneity regime [16]. We propose that Eq. (14) may provide a reasonable definition of the equilibrium superfluid density $\rho_s(\mathbf{r})$ in the weakly inhomogeneous case, when we take its static limit and the $\mathbf{k}=0$ component of its Fourier transform with respect to $\mathbf{r}_1 - \mathbf{r}_2$. The functional derivative entering the RHS of Eq. (14) is a five-point correlation function, to be evaluated in this case on the equilibrium state.

With the above definition of the superfluid density, we first proceed to extend Landau's hydrodynamic theory for the homogeneous superfluid to finite-frequency phenomena. We use for this purpose the well-known memory function formalism, as described *e.g.* in the book of Forster [19]. We assume isothermal conditions, *i.e.* we neglect the couplings between temperature and density fluctuations. The form of the generalized hydrodynamic equations is dictated by some general considerations: (i) invariance under a Galileian transformation and Onsager symmetry must hold; (ii) as a consequence of the zero-force and zero-torque theorems, the time derivative of the current density $\mathbf{j}$ is driven by the divergence of a symmetric tensor of the second rank; (iii) the time derivative of the superfluid velocity $\mathbf{u}_s$ is the gradient of a scalar, in view of its irrotational character below threshold for vortex generation; and (iv) the internal driving forces are determined by the divergence of the normal-fluid velocity $\mathbf{u}_n$ and of the interdiffusion current $\mathbf{j}_r \equiv \mathbf{j} - n\mathbf{u}_n = \rho_s(\mathbf{u}_s - \mathbf{u}_n)$.

The generalized hydrodynamic approach is easily extended to a weakly inhomogeneous superfluid in isothermal conditions [16]. The general properties that we have listed above to infer the structure of the generalized hydrodynamic equations in the homogeneous case remain valid. In addition, a Ward identity is essential in relating the effects of a weak inhomogeneity on the excess kernels to their density dependence: it allows the inhomogeneous kernels to be set equal to those of the homogeneous fluid at the local densities of superfluid and normal fluid. The resulting dynamical equations are

$$-im\omega\delta\mathbf{j}(\mathbf{r},\omega) = n(\mathbf{r})[-i\omega\mathbf{a}(\mathbf{r},\omega) - \vec{\nabla}(\delta n(\mathbf{r},\omega)/n^2 K_T)] + \vec{\nabla}\cdot\overset{\leftrightarrow}{\sigma}(\mathbf{r},\omega) \tag{16}$$

and

$$-im\omega\delta\mathbf{u}_s(\mathbf{r},\omega) = -i\omega\mathbf{a}_s(\mathbf{r},\omega)$$
$$- \vec{\nabla}[(n^2 K_T)^{-1}\delta n(\mathbf{r},\omega) - (Ts/c_V)\delta s(\mathbf{r},\omega)] + \vec{\nabla}\cdot\overset{\leftrightarrow}{\sigma}{}^{(s)}(\mathbf{r},\omega) \quad , \tag{17}$$

where $s$ and $c_V$ are the entropy and heat capacity per particle, and the stress tensors are given by

$$\sigma_{ij} = [\eta(\omega) - p_0(n)/i\omega]\left(\frac{\partial u_{ni}}{\partial r_j} + \frac{\partial u_{nj}}{\partial r_i} - \frac{2}{3}\delta_{ij}\vec{\nabla}\cdot\mathbf{u}_n\right)$$
$$+ \delta_{ij}[\zeta_2(\omega)\vec{\nabla}\cdot\mathbf{u}_n + \zeta_1(\omega)\vec{\nabla}\cdot\mathbf{j}_r] \tag{18}$$

and

$$\sigma_{ij}^{(s)} = \delta_{ij}[\zeta_3(\omega)\vec{\nabla}\cdot\mathbf{j}_r + \zeta_4(\omega)\vec{\nabla}\cdot\mathbf{u}_n] \quad, \tag{19}$$

$p_0(n)$ being the ideal-gas pressure. The ALDA contributions have been explicitly written in Eqs. (16) and (17) in terms of the total particle density fluctuations and of the entropy fluctuations, the latter being associated with fluctuations in the density of the non-condensate. The corresponding thermodynamic coefficients are to be evaluated on the homogeneous fluid at the local equilibrium density. Evidently, the ALDA driving forces are responsible for the first and second sound modes.

The visco-elastic coefficients in Eqs. (18) and (19) are related to a set of excess kernels according to the following equations:

$$\zeta_1(\omega) = -(i\omega)^{-1}[f_{j_L u_s}(\omega) - \partial p_{xc}/\partial n|_T] \quad, \tag{20}$$

$$\zeta_2(\omega) = -(i\omega)^{-1}[f_{j_L j_L}(\omega) - 4f_{j_T j_T}(\omega)/3 - n\partial p_{xc}/\partial n|_T] \quad, \tag{21}$$

$$\zeta_3(\omega) = -(i\omega)^{-1}[f_{u_s u_s}(\omega) - \partial\mu_{xc}/\partial n|_T - (Ts/c_V)\partial\mu_{xc}/\partial T|_n] \quad, \tag{22}$$

$$\zeta_4(\omega) = -(i\omega)^{-1}[f_{u_s j_L}(\omega) - \partial p_{xc}/\partial n|_T] \tag{23}$$

and

$$\eta(\omega) = -(i\omega)^{-1}f_{j_T j_T}(\omega) \quad, \tag{24}$$

with $\zeta_1 = \zeta_4$ from Onsager symmetry. As already remarked, the kernels in these equations are those of the homogeneous fluid at the local equilibrium densities of superfluid and normal fluid at temperature $T$. The xc pressure and chemical potential are similarly those of the homogeneous fluid at the local equilibrium density.

Within the memory function formalism, the visco-elastic spectra in Eqs. (18) and (19) can be calculated from the linear response functions of the homogeneous superfluid through finite-frequency forms of the Kubo relations. These will be used in § 4.2 for a Bose vapour.

## 4. Evaluation of dynamic kernels

### 4.1. THE ELECTRON FLUID IN THE NORMAL STATE

As we have already remarked at the end of § 2, Eq. (8) implies that the visco-elastic coefficients are related to the current-current response function of the homogeneous electron fluid by a finite-frequency extension of the Kubo relations. Precisely,

$$\text{Im } f_{xc}^{L(T)}(\omega) = \lim_{k\to 0}\frac{m^2\omega^2}{n^2 k^2}\text{Im }\tilde{\chi}_{L(T)}(k,\omega) \quad, \tag{25}$$

where $\tilde{\chi}_{L(T)}(k,\omega)$ is the proper longitudinal (or transverse) current-current response [13]. This emphasizes that the plasmon does not contribute to the xc kernels.

Starting from the equation of motion for the current density fluctuation operator $\hat{j}_{\vec{q}i}$, the following exact long-wavelength result can be derived:

$$\mathrm{Im}\,\tilde{\chi}_{ij}(\vec{k},\omega) = -(m^2\omega^4)^{-1} \sum_{\vec{q},\vec{q}'} \sum_{s,s'} \Gamma_{is}(\vec{q},\vec{k}) \Gamma_{js'}(\vec{q}',-\vec{k})$$

$$\mathrm{x}\,\mathrm{Im}\{i\int_0^\infty dt\,\exp[i(\omega+i\eta)t] < [\hat{j}_{\vec{q}s}\hat{\rho}_{-\vec{q}},\hat{j}_{\vec{q}'s'}\hat{\rho}_{-\vec{q}'}]>\} + o(k^2)\ ,\quad (26)$$

$\hat{\rho}_{\vec{q}}$ being the density fluctuation operator and the detailed expression of the coefficients $\Gamma$ being given in ref. [15]. Decoupling of the four-point correlation function entering the RHS of Eq. (26) into all possible products of two-point correlation functions leads to the following approximate result [13]:

$$\mathrm{Im}\,f_{xc}^{L(T)}(\omega) = -\int_0^\infty \frac{d\omega'}{\pi} \int \frac{d^3q}{(2\pi)^3} \frac{(4\pi e^2/n\omega')^2}{(\omega-\omega')^2} \mathrm{Im}\,\chi_L(q,\omega-\omega')$$

$$\mathrm{x}\left[a_{L(T)}\,\mathrm{Im}\,\chi_L(q,\omega') + b_{L(T)}(\omega'/\omega)^2\,\mathrm{Im}\,\chi_T(q,\omega')\right]\ .\quad (27)$$

Here, the $a$'s and $b$'s are numerical coefficients given by $a_L = 23 a_T/16 = 23/30$ and $b_L = 4 b_T/3 = 8/15$. Equation (27) only includes the contribution from excitations of two correlated electron-hole pairs. It can also be derived by means of second-order perturbation theory followed by approximate resummation to infinite order through renormalization of single-particle excitations from bare to screened [20]. For further details of this calculation, including the various possible choices for the response functions in the RHS of Eq. (26) and the inclusion of final-state exchange processes restoring the exactly known coefficients of the $\omega^{-3/2}$ decay of the spectra at high frequency [21], the reader is referred to the original papers [13, 15].

Equation (27) shows that both the longitudinal and the transverse spectrum involve an LL and an LT channel, the former becoming dominant at high frequency and the latter at low frequency (on a frequency scale set by the plasma frequency $\omega_{pl}$). The values of the numerical coefficients in Eq. (27) are such that $\mathrm{Im}\,f_{xc}^T = (16/23)\,\mathrm{Im}\,f_{xc}^L$ at high frequency and $\mathrm{Im}\,f_{xc}^T = (3/4)\,\mathrm{Im}\,f_{xc}^L$ at low frequency: as a consequence, the two spectra have very similar shapes, the relation $\mathrm{Im}\,f_{xc}^T \cong 0.72\,\mathrm{Im}\,f_{xc}^L$ being approximately valid at all frequencies. This is shown in Figure 1, which reports the two spectra for the electron fluid at a value of the coupling strength $r_s = 3$ in comparison with an earlier attempt by Gross and Kohn [22] at estimating the longitudinal spectrum by a smooth interpolation between high and low frequency. In contrast, our microscopic calculations show a sharp spectral threshold at $2\omega_{pl}$, which is due to the opening of a two-plasmon channel in the two-pair excitation spectrum [23]. The large spectral strength of the plasmon excitation at long wavelengths, as compared to single-pair excitations, accumulates most of the oscillator strength for two-pair processes in the spectral region just above $2\omega_{pl}$.

The real part of the longitudinal and transverse kernels is obtained from the spectra illustrated in Figure 1 by means of Kramers-Kronig relations, *i.e. via* Hilbert transform and fulfilment of the third-moment sum rules. As a consequence of the spectral threshold at $2\omega_{pl}$, a sharp minimum is found in $\mathrm{Re}\,f_{xc}^{L(T)}$ at this frequency. Strong xc effects are present in $\mathrm{Re}\,f_{xc}^{L(T)}$ in the region of the plasma frequency, where these functions decrease with increasing $\omega$ towards their minimum at $2\omega_{pl}$. This theoretical result is in accord with the available data on the plasmon dispersion coefficient from Electron Energy Loss experiments on the alkali metals [24], although band-structure effects will also play a significant role in a fully quantitative comparison between theory and experiment.

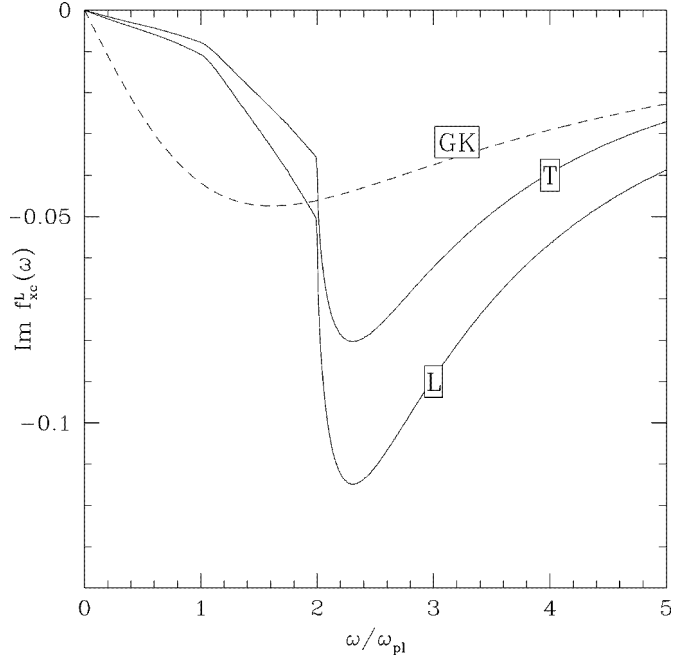

*Figure 1.* $\mathrm{Im} f_{xc}^{L,T}(\omega)$ for the electron gas at $r_s = 3$ in units of $2\omega_{pl}/n$, as functions of $\omega/\omega_{pl}$. The dashed line gives the Gross-Kohn interpolation scheme.

As a last point we briefly record the main results that have been obtained by this approach [15] for the elastic moduli and the viscosity coefficients of the fully degenerate electron gas in dimensionalities D = 3 and D = 2. These are: (i) the calculated electron-gas compressibility is in good agreement with the available Quantum Monte Carlo data on the xc energy; (ii) xc effects decrease the shear elastic constant for $r_s \geq 6$ in D = 3 and at all $r_s$ in D = 2; (iii) the bulk viscosity vanishes identically at zero temperature; and (iv) the shear viscosity takes a finite value on inclusion of dynamic xc effects and this value rapidly decreases with increasing $r_s$.

4.2. THE DILUTE BOSON SUPERFLUID

The visco-elastic spectra generalizing sound wave attenuation and shear viscosity to finite frequency are related by Eqs. (16) and (18) to the longitudinal and transverse current-current response functions of the homogeneous fluid. Precisely, these relations are

$$\mathrm{Re}[\zeta_2(\omega) + \frac{4}{3}\eta(\omega)] = \lim_{k \to 0} \frac{-\omega m^2}{k^2} \mathrm{Im}\chi_{j_L j_L}(k,\omega) \qquad (28)$$

and

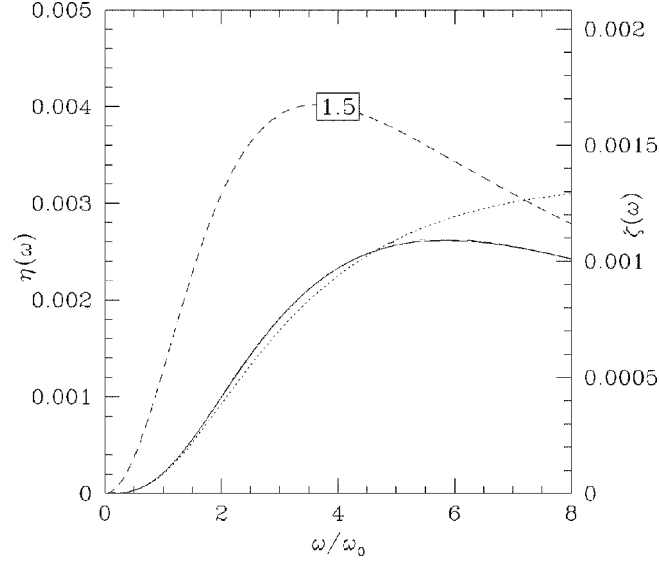

*Figure 2.* Spectra of shear viscosity (left scale) and bulk viscosity (right scale) as functions of $\omega/\omega_0$ at $T = 0$ (full curve) and $T = 1.5\omega_0$ (dashes). The dots show the result obtained at $T = 0$ without a cutoff in the potential.

$$\text{Re } \eta(\omega) = \lim_{k \to 0} \frac{-\omega m^2}{k^2} \text{Im } \chi_{j_T j_T}(k, \omega) \quad . \tag{29}$$

Of course, the imaginary parts of $\zeta_2(\omega)$ and $\eta(\omega)$, which are finite-frequency bulk and shear elastic moduli, are related to the real parts by Kramers-Kronig relations. Equations (28) and (29) are evidently analogous to Eq. (25) and can be evaluated in the same approach which led to Eq. (27) for the electron fluid. For the boson superfluid this approach may also be viewed as amounting to an expansion of the single-particle self-energy to second order in perturbation theory, followed by renormalization of the single-particle propagators into Bogolubov phonons. The extension of the theory to finite temperature is easily achieved, by introducing in the calculation of the response functions the thermal occupation factors $B(\omega) = [\exp(\beta\omega) - 1]^{-1}$ as required by the fluctuation-dissipation theorem.

The case of present interest concerns the confined condensates of Na or Rb atoms inside magnetic traps, in which the atomic interactions can be modelled by a repulsive Fermi model potential involving an *s*-wave scattering length $a$ as a single parameter. To regularize the interatomic potential $v_k$ at large momenta **k**, we introduce a cut-off by writing

$$v_k = (4\pi a/m)\exp(-k^2/2\sigma^2) \quad . \tag{29}$$

The corresponding Bogolubov spectrum is

$$\omega_k = \left[(nv_k k^2/m) + (k^2/2m)^2\right]^{1/2} \quad . \tag{30}$$

A lengthy but straightforward calculation, whose details will be reported elsewhere [25], yields for the shear and bulk viscosity spectra in the collisionless regime the result

$$\text{Re}[\eta(\omega)] = \frac{12}{5}\text{Re}[\zeta_2(\omega)] = \frac{v_k^2 k^6}{240\pi m^2 \omega_k^3 \omega_k'}(1 - \frac{2}{7}\alpha_k + \frac{3}{35}\alpha_k^2)\frac{\sinh(\beta\omega_k)}{\cosh(\beta\omega_k)-1} \quad . \quad (31)$$

Here, $k$ is defined by $2\omega_k = \omega$, $\omega_k' = d\omega_k/dk$ and $\alpha_k = (\mathbf{k}\cdot\overset{\perp}{\nabla}v_k)/v_k$. At low frequency we find $\eta(\omega) \propto \omega^3$ if $\beta\omega \gg 1$ and $\eta(\omega) \propto \omega^2 T$ if $\beta\omega \ll 1$.

Figure 2 shows these visco-elastic spectra as functions of frequency at T = 0 and at a higher temperature on a scale set by the frequency $\omega_0 = 4\pi na/m$. A further result is obtained by evaluating the longitudinal spectrum at the sound-wave frequency $ck$, whereby one obtains the width $\Delta(k)$ of the phonon as

$$\Delta(k) = 3k^5/(640\pi mn) \quad (32)$$

at T = 0 and

$$\Delta(k) = 3\pi^3 k(k_B T)^4/(40mc^4) \quad (33)$$

for $\beta\omega \ll 1$. These results for the phonon linewidth are well known from earlier work of Belyaev and Popov [26].

It can be shown that the results reported above for the visco-elastic current spectra at T = 0 correspond to the so-called one-loop approximation introduced in the context of Bose gases by Wong and Gould [27]. We have calculated the other spectra of relevance to Eqs. (16) - (19) by the same method, starting from the finite-frequency Kubo relations

$$\text{Re}[\zeta_3(\omega)] = \lim_{k\to 0}\frac{-\omega}{k^2}\text{Im}\,\chi_{u_s u_s}(k,\omega) \quad (34)$$

and

$$\text{Re}[\zeta_1(\omega)] = \text{Re}[\zeta_4(\omega)] = \lim_{k\to 0}\frac{-\omega m}{k^2}\text{Im}\,\chi_{j_L u_s}(k,\omega) \quad . \quad (35)$$

It turns out that within the one-loop approximation all these dissipation spectra can be expressed in terms of four exchange-correlation building blocks: a proper condensate kernel related to the shift in local chemical potential due to interactions in the fluid away from equilibrium, two noncondensate kernels (the irreducible proper parts of the longitudinal and transverse current-current response) and a cross condensate-noncondensate vertex function. Actual calculation of these spectral functions at T = 0 shows that in a collisionless regime they take the same value aside from simple multiplying factors. The results are

$$\frac{12}{5}\text{Re}[\zeta_2(\omega)] = \frac{2}{5}\text{Re}[mn\,\zeta_1(\omega)] = \frac{8}{15}\text{Re}[m^2 n^2 \zeta_3(\omega)] = \text{Re}[\eta(\omega)] \quad . \quad (36)$$

## 5. Summary and concluding remarks

We have reviewed the progress recently made in developing the theory of the dynamics of weakly inhomogeneous quantal fluids in the linear response regime from knowledge of the dynamics of the corresponding homogeneous fluid. The present focus in this area is on the degenerate electron gas and on the dilute Bose superfluid, for which the basic theoretical concepts and some inputs for applications have been developed. The fruitfulness of this approach remains to be tested through applications to specific physical problems. Such applications are already under way in the case of electron fluids [28].

**Acknowledgments**

This work is supported by the Istituto Nazionale di Fisica della Materia through the Advanced Research Project on BEC. We acknowledge very fruitful collaborations with Dr S. Conti.